\documentclass{webofc}
\usepackage[varg]{txfonts} 
\usepackage{graphicx} 
\usepackage{epstopdf} 
\begin{document}
\title{Another look at the Landau gauge three-gluon vertex}
%
%

\author{\firstname{Guilherme T. R.} \lastname{Catumba}\inst{1,2}\fnsep\thanks{\email{gtelo@ific.uv.es}} \and
        \firstname{Orlando} \lastname{Oliveira}\inst{2}\fnsep\thanks{\email{orlando@uc.pt}} \and
        \firstname{Paulo J.} \lastname{Silva}\inst{2}\fnsep\thanks{Speaker, \email{psilva@uc.pt}}
}

\institute{IFIC - University of Valencia - Spain
\and
       CFisUC, Department of Physics, University of Coimbra, 3004-516 Coimbra, Portugal   
          }

\abstract{%
We revisit the computation of the three-gluon vertex in the Landau gauge
using lattice QCD simulations with large physical volumes of $\sim$ (6.5
fm)$^ 4$  and $\sim$ (8 fm)$^ 4$ and large statistical ensembles.
For the kinematical configuration analysed, that is described by a unique
form factor, an evaluation of the lattice artefacts is also
performed. Particular attention is given to the low energy behaviour of
vertex and its connection with evidence (or lack of it)
of infrared ghost dominance.
}
\maketitle

\section{Introduction}

The amputated three-gluon correlation function, herein also named vertex,
is a fundamental QCD correlation function that allows for the
extraction of a strong coupling constant and the computation of a static
potential between color charges. Our aim is to revisit a previous
calculation \cite{duarte2016, proc2016} using an improved statistics
for the larger volume and estimate the corresponding lattice artefacts.

Lattice simulations measure the correlation function
$G^{a_1 a_2 a_3}_{\mu_1 \mu_2 \mu_3} (p_1, p_2, p_3)$ via the computation of
\begin{equation}
   \langle A^{a_1}_{\mu_1} (p_1) \, A^{a_2}_{\mu_2} (p_2) \, A^{a_3}_{\mu_3} (p_3) \rangle =   V \, \delta( p_1 + p_2 + p_3) ~
   {G^{a_1 a_2 a_3}_{\mu_1 \mu_2 \mu_3} (p_1, p_2, p_3)} \ .
\end{equation}
In terms of the gluon propagator $D$ and of the
one-particle irreducible three-gluon diagram  (1PI) $\Gamma$ the
correlation function is given by
\begin{equation}
  {G^{a_1a_2a_3}_{\mu_1\mu_2\mu_3} (p_1, p_2, p_3)}  =   D^{a_1b_1}_{\mu_1\nu_1}(p_1) ~ D^{a_2b_2}_{\mu_2\nu_2}(p_2) ~ D^{a_3b_3}_{\mu_3\nu_3}(p_3) ~
    {\Gamma^{b_1b_2b_3}_{\nu_1\nu_2\nu_3} (p_1, p_2, p_3)} .
\end{equation}
The three-gluon vertex reads
\begin{equation}
 \Gamma^{a_1 a_2 a_3}_{\mu_1 \mu_2 \mu_3} (p_1,  p_2, p_3) = f_{a_1 a_2 a_3} \Gamma_{\mu_1 \mu_2 \mu_3} (p_1, p_2, p_3)
\end{equation}
and given that Bose symmetry requires the 1PI function to be symmetric
under permutations of any pair $(p_i, a_i, \mu_i)$ it follows that
$\Gamma_{\mu_1 \mu_2 \mu_3} (p_1, p_2, p_3)$
has to  be antisymmetric under the interchange of any pair
$(p_i, \mu_i)$. The description of
$\Gamma_{\mu_1 \mu_2 \mu_3} (p_1, p_2, p_3)$ in the continuum
formulation requires six Lorentz invariant form factors. Two of the
form factors are associated with the transverse component of the
vertex, while the remaining define its longitudinal component \cite{ballchiu}.

Here we consider the evaluation of the vertex for
the asymmetric momentum configuration that is defined by
$p_2=0$, as in \cite{duarte2016, proc2016, alles}.
Then, the three point connected Green function is given by
\begin{equation}
   G_{\mu_1\mu_2\mu_3} (p, 0, -p)  =    V \frac{N_c(N^2_c-1)}{4}  \left[D(p^2)\right]^2 \, D(0) \frac{\Gamma (p^2)}{3} ~ ~ p_{\mu_2} ~T_{\mu_1\mu_3} (p)
\end{equation}
and, therefore,
\begin{equation}
    G_{\mu \, \alpha \,\mu} (p, 0, -p) \, p_\alpha = V \frac{N_c(N^2_c-1)}{4}
   \, \left[D(p^2)\right]^2 \, D(0) ~~\Gamma (p^2) ~~ p^2 .
\end{equation}
It follows that the form factor  $\Gamma (p^2)$ can be measured by
computing the ratio
\begin{equation}
 G_{\mu \alpha \mu} (p, 0, -p) p_\alpha  / \left[D(p^2)\right]^2 \,
 D(0) \ .
\end{equation}
For large momentum, the gluon propagator becomes small and its
fluctuations induce  large variations for $\Gamma (p^2)$ that prevent
a precise measurement of this form factor in the UV regime. Indeed,
such large fluctuations have been observed in previous simulations;
see e.g. \cite{duarte2016}
and references therein.

\section{The lattice setup}

In the following we show results from a simulation using a $64^4$
lattice with an ensemble of 2000 configurations already studied in
\cite{duarte2016},
together with a simulation using a $80^4$ lattice and an ensemble of
1800 configurations.
The ensembles were generated with the Wilson gauge action at
$\beta=6.0$ using Chroma library \cite{chroma}.
The gauge configurations have been rotated to the Landau gauge using the
Fourier accelerated Steepest Descent method \cite{davies}. Fast Fourier
transforms were implemented using the  PFFT library \cite{pfft}.
Our definition for the gluon field being
\begin{equation}
a g_0 A_\mu (x + a \hat{e}_\mu)  = \frac{ U_\mu (x) - U^\dagger (x)}{ 2 i g_0} 
      - \frac{\mbox{Tr} \left[ U_\mu (x) - U^\dagger (x) \right]}{6 i g_0} 
\end{equation}
with its corresponding field in momentum space being given by
\begin{equation}
A_\mu (\hat{p}) = \sum_x e^{- i q (x + a \hat{e}_\mu) } \, A_\mu (x + a \hat{e}_\mu) \,\,,\,\, q_\mu = \frac{2 \, \pi \, n_\mu}{a \, L_\mu}.
\end{equation}
An improved definition for lattice momenta, motivated by lattice perturbation theory, is 
\begin{equation}
  p_{\mu}=\frac{2}{a} \sin\left(\frac{\pi n_{\mu}}{L_{\mu}}\right).
\end{equation}

\section{Handling of noise and lattice artefacts}

We use two approaches to handle the problem of the large statistical
fluctuations at high momenta already discussed in
\cite{Catumba:2021hcx,guitese}, namely

\begin{itemize}
\item explore the ambiguity on the scale setting that allow us to bin
  the momentum data --- the momentum are grouped in bins and for each bin the
  lattice data is replaced by its weighted average, using as weight the inverse
  of the statistical error of the bin data points;
\item perform a $H(4)$ extrapolation of the lattice data
  \cite{becirevic1999, soto2009} --- such procedure uses the naive lattice momentum $q_{\mu}$
  and explores the lattice   $H(4)$ symmetry group that is associated with an
  hypercube. Accordingly, any lattice scalar function $F$ depends on the $H(4)$ invariants
\begin{displaymath}
  q^2 = q^{[2]} = \sum_\mu q^2_\mu ,  \quad
  q^{[4]} = \sum_\mu q^4_\mu ,  \quad
  q^{[6]} = \sum_\mu q^6_\mu ,  \quad
  q^{[8]} = \sum_\mu q^8_\mu ,
\end{displaymath}
and one writes $F_{Lat} = F(q^{[2]}, q^{[4]}, q^{[6]}, q^{[8]})$. The
corresponding  function in the continuum limit  is given by
$F = F(q^{[2]}, 0, 0, 0)$ up to corrections $\mathcal{O}(a^2)$.
The procedure requires several data points with the same $q^2$ but
different $q^{[4]}$, $q^{[6]}$, $q^{[8]}$. $F$ can be computed via an
extrapolation of $F_{Lat}$ if one assumes that $F_{Lat}$ can be
written as a power series of the H(4) invariants. Herein, we ignore
contributions due to $q^{[6]}$ and $q^{[8]}$ and consider a linear
extrapolation in $q^{[4]}$.


\end{itemize}

\section{Results}

\begin{figure*}
\centering
\includegraphics[width=0.47\textwidth]{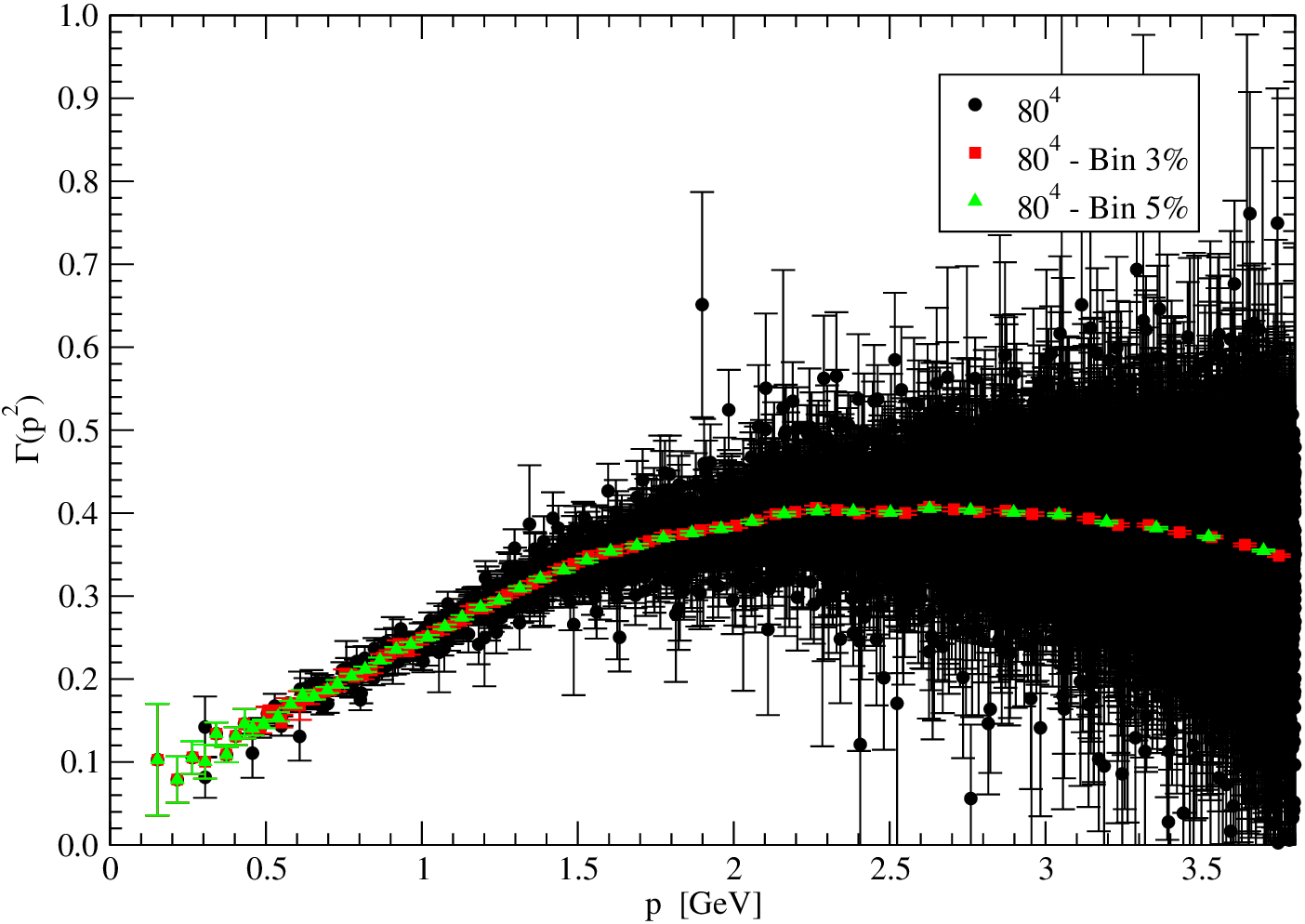}\hspace*{0.5cm}
\includegraphics[width=0.47\textwidth]{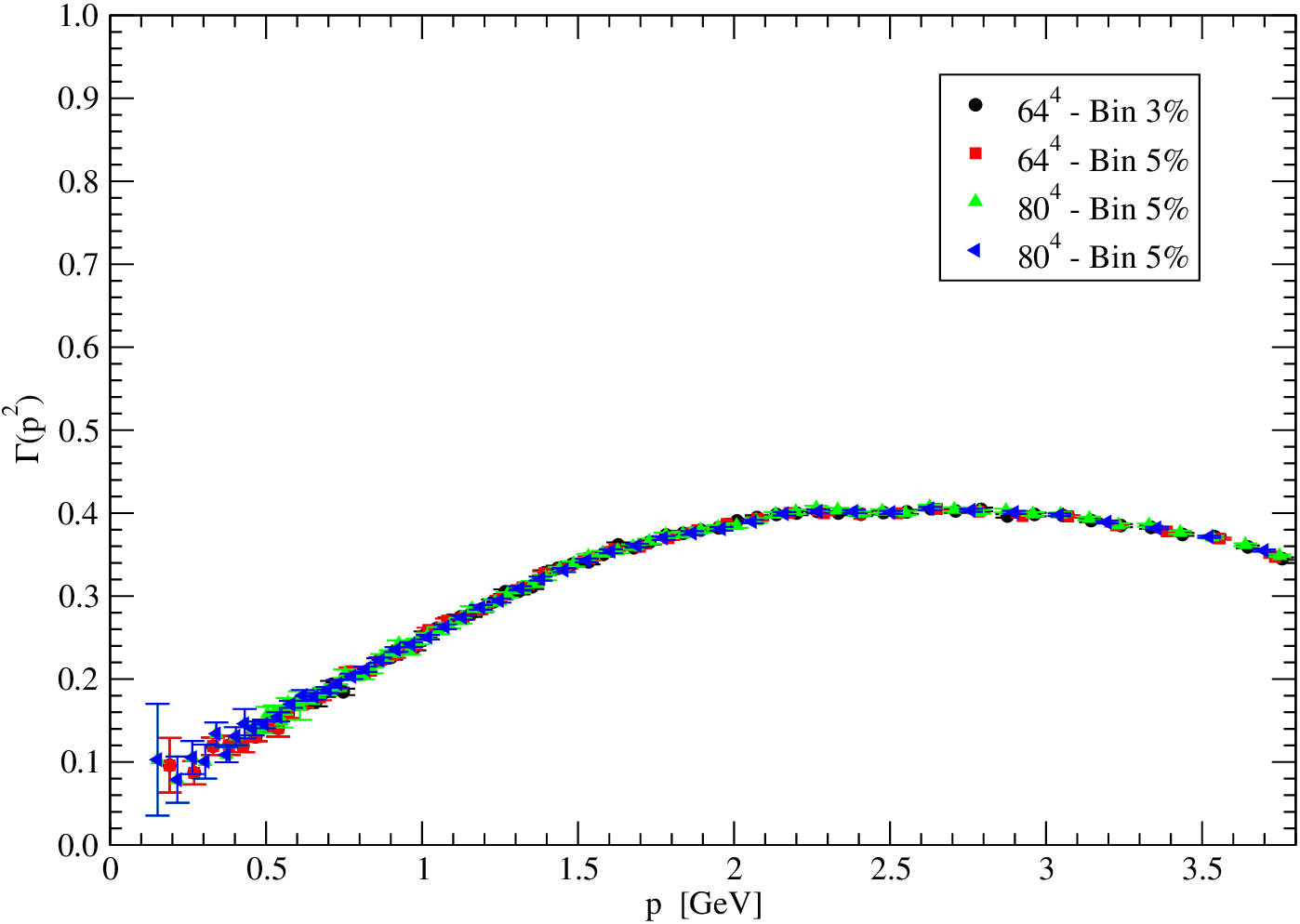}
\caption{Left plot: original and binned data for $\Gamma (p^2)$, $80^4$ lattice. Right plot: Comparison of binned data for  $\Gamma (p^2)$ and both lattice volumes.  }
\label{binned}    
\end{figure*}

The data for $\Gamma (p^2)$ both binned and unbinned is reported in the left plot of Figure \ref{binned} for the largest lattice. Clearly, by binning the data the large
statistical errors in the high momentum region are suppressed
resulting in a well defined and smooth curve. On the right plot of Figure \ref{binned}, the binned data for the two lattice
volumes is compared. The good agreement within errors of both sets of
data suggest that the finite volume effects are under control and are small.

\begin{figure*}
\centering
\includegraphics[width=0.47\textwidth]{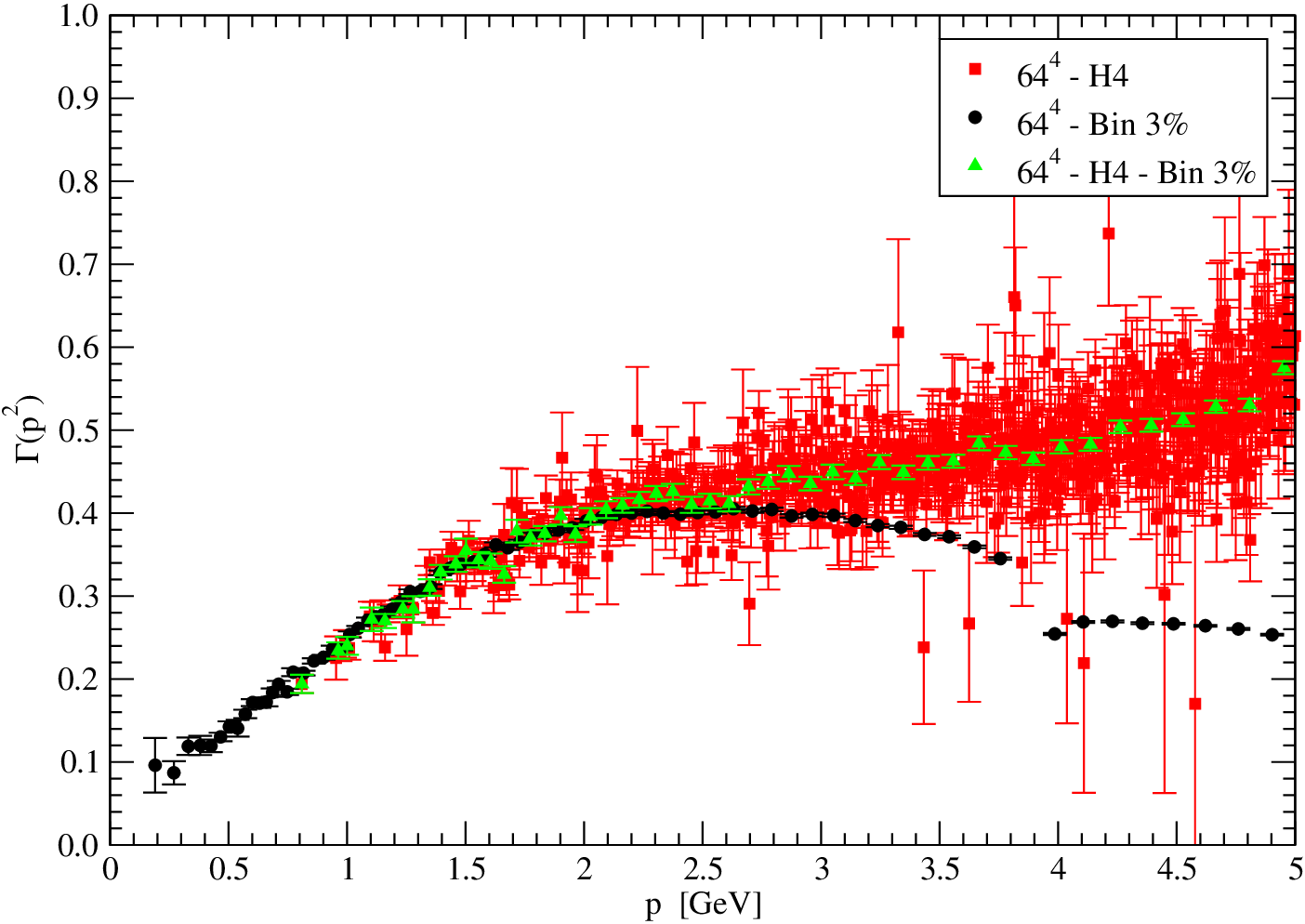}\hspace*{0.5cm}
\includegraphics[width=0.47\textwidth]{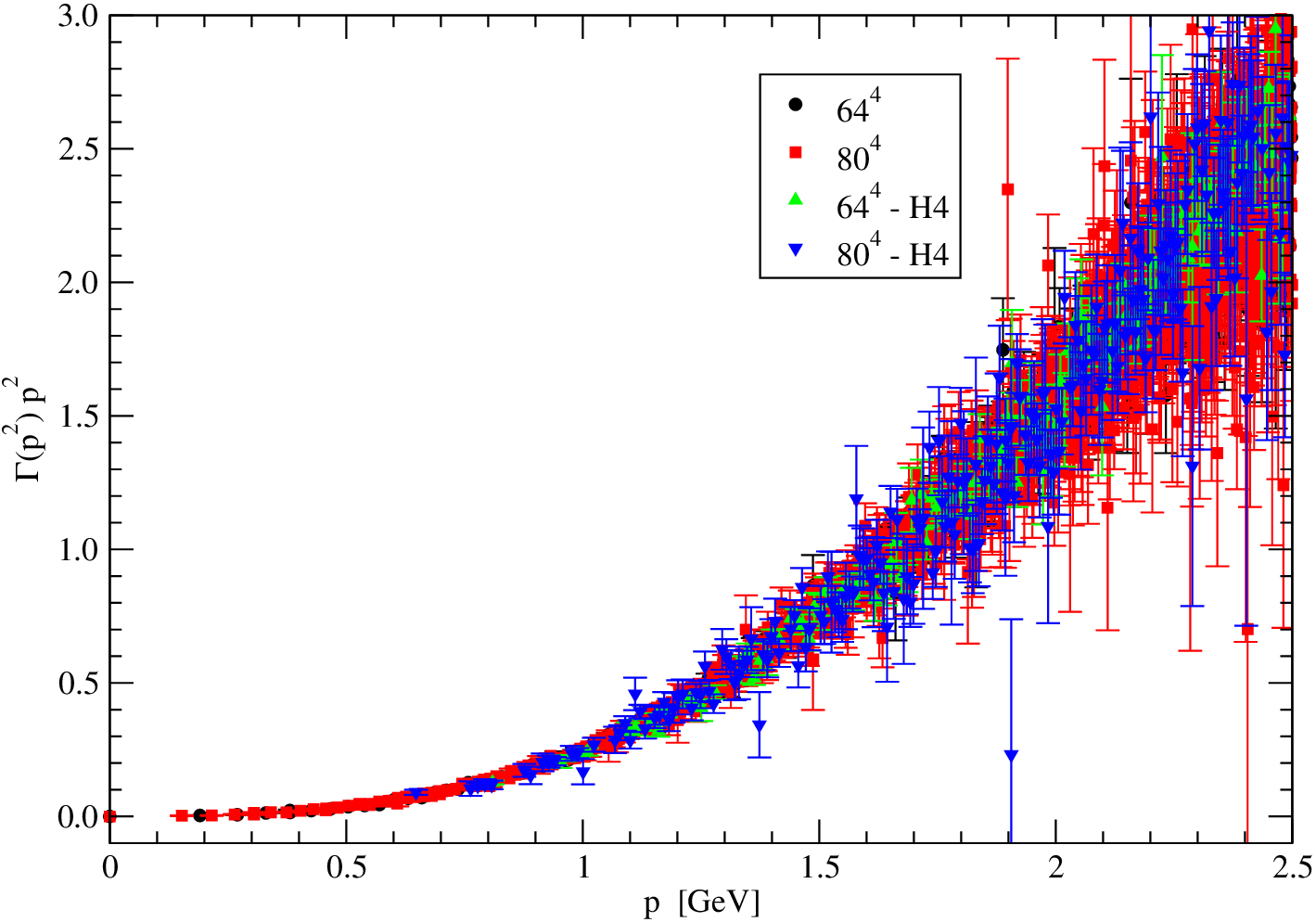}
\caption{Left plot: results of the H(4) extrapolation of  $\Gamma (p^2)$ for the
  $64^4$ lattice volume. Right plot: original and H(4) data for $p^2 \Gamma(p^2)$ and for both lattice volumes in the infrared region.}
\label{H4extr}
\end{figure*}

The linear H(4) extrapolation of the $64^4$ lattice data, including the binned data, is compared with the original binned data in the left plot of Figure \ref{H4extr}. Up to momentum
$p \sim 2.5$ GeV both the H(4) extrapolation and the binned data are in good
agreement. However, for large $p$ the H(4) extrapolation overestimates
$\Gamma$ in comparison with the binned lattice data. As seen in the right plot of Figure
\ref{H4extr}, in the low momentum regime the two sets of data agree
within errors.

As can be observed in all the Figures the form factor does not seem to
change sign in the infrared region. Recall that the change of sign is, in general,
understood as an indication of ghost dominance in the infrared.

\section{Infrared behaviour of $\Gamma(p^2)$}

The lack of observing a change of sign in the form factor $\Gamma(p^2)$ can be an
indication that either it does not happen and there is no ghost dominance
in the infrared region or the change of sign occurs at momenta that
are smaller than those accessed in the simulations. Here, we explore
further the infrared behavior of $\Gamma(p^2)$ by fitting the lattice
data to various functional forms. In the following, only the $80^4$
lattice data for momenta below 1GeV will be considered.

\begin{figure*} 
  \centering
  \includegraphics[width=0.47\textwidth]{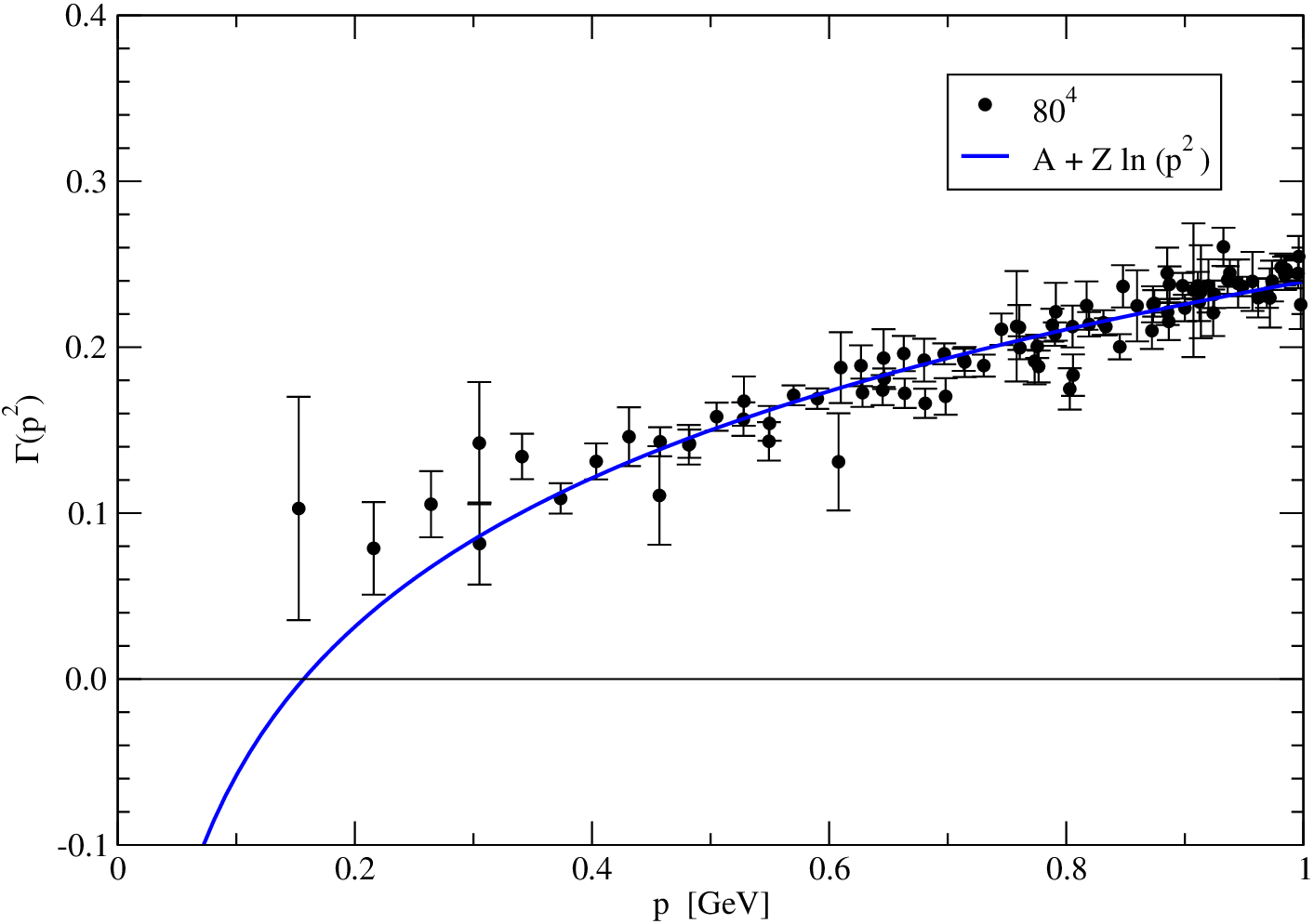}\hspace*{0.5cm}
  \includegraphics[width=0.47\textwidth]{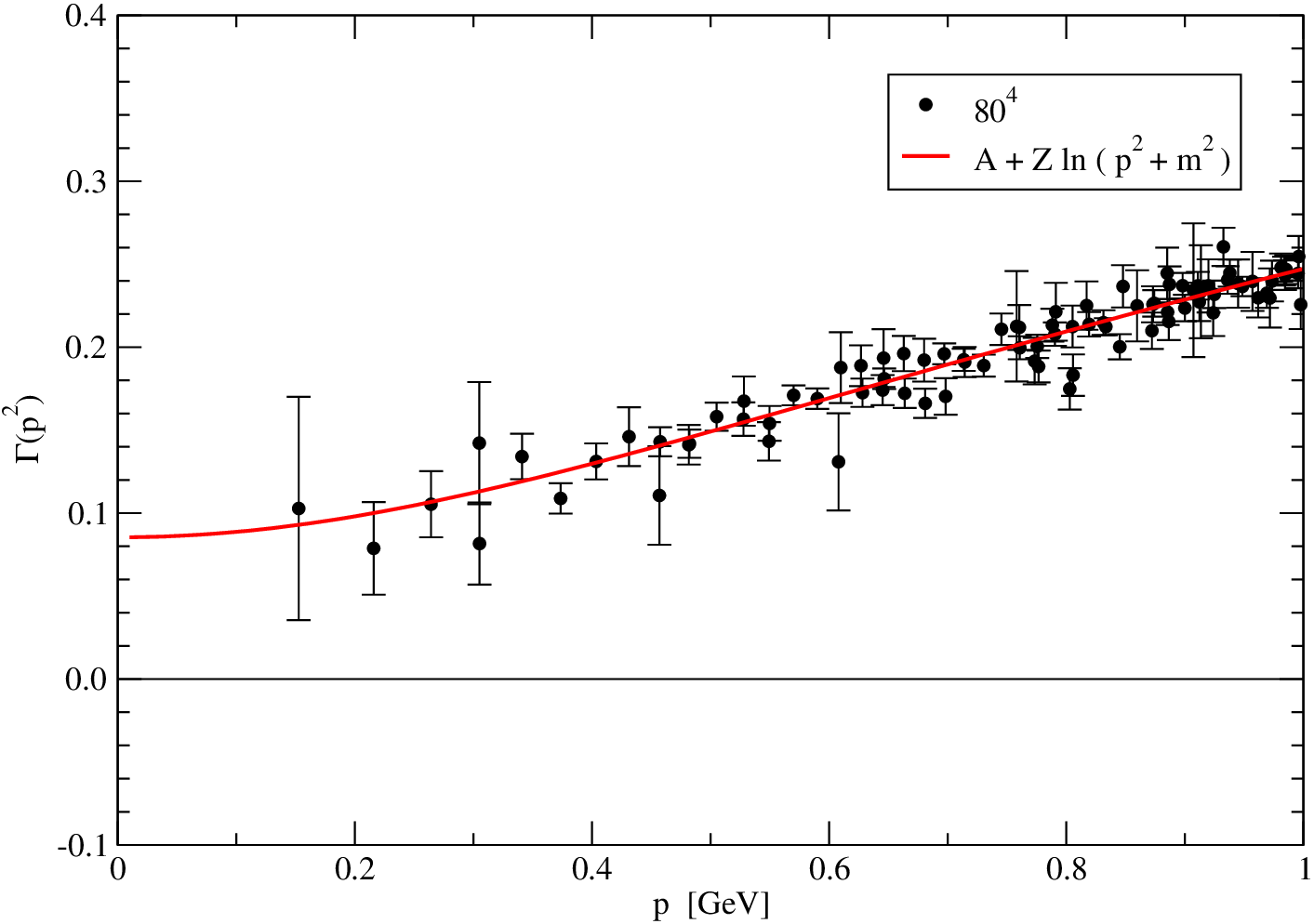} 
  \caption{Infrared $80^4$ lattice data for $\Gamma(p^2)$ together with some fitting functions. Left plot: $\Gamma (p^2) = A + Z \ln(p^2)$; right plot: $\Gamma (p^2) = A + Z \ln(p^2+m^2)$.  }
  \label{zerocrossing1}
\end{figure*}

The best fits of the lattice data to $\Gamma_1(p^2)=A + Z \ln(p^2)$ and
$\Gamma_2(p^2)=A + Z \ln(p^2+m^2)$
are reported in Figure \ref{zerocrossing1}. $\Gamma_1(p^2)$ was used in recent studies of the infrared
three-gluon vertex, see \cite{guitese} for details, and 
$\Gamma_2(p^2)$ is a regularized version of $\Gamma_1(p^2)$ that is,
in principle, finite over the full range of $p^2$.
The best fit to $\Gamma_1(p^2)$ results in a $\chi^2/d.o.f. = 1.23$,
$A=0.2395(16)$ and  $Z=0.0646(21)$. Hence, the zero crossing
should occur at  $p_o=157$MeV. The best fit for $\Gamma_2(p^2)$
gives $A=0.208(24)$, $Z=0.124(27)$ and
$m=0.61(15)$ GeV, with a $\chi^2/d.o.f. = 0.95$. No change of sign
 can be associated with this function; see the right plot of Figure \ref{zerocrossing1}.

\begin{figure*} 
  \centering
  \includegraphics[width=0.47\textwidth]{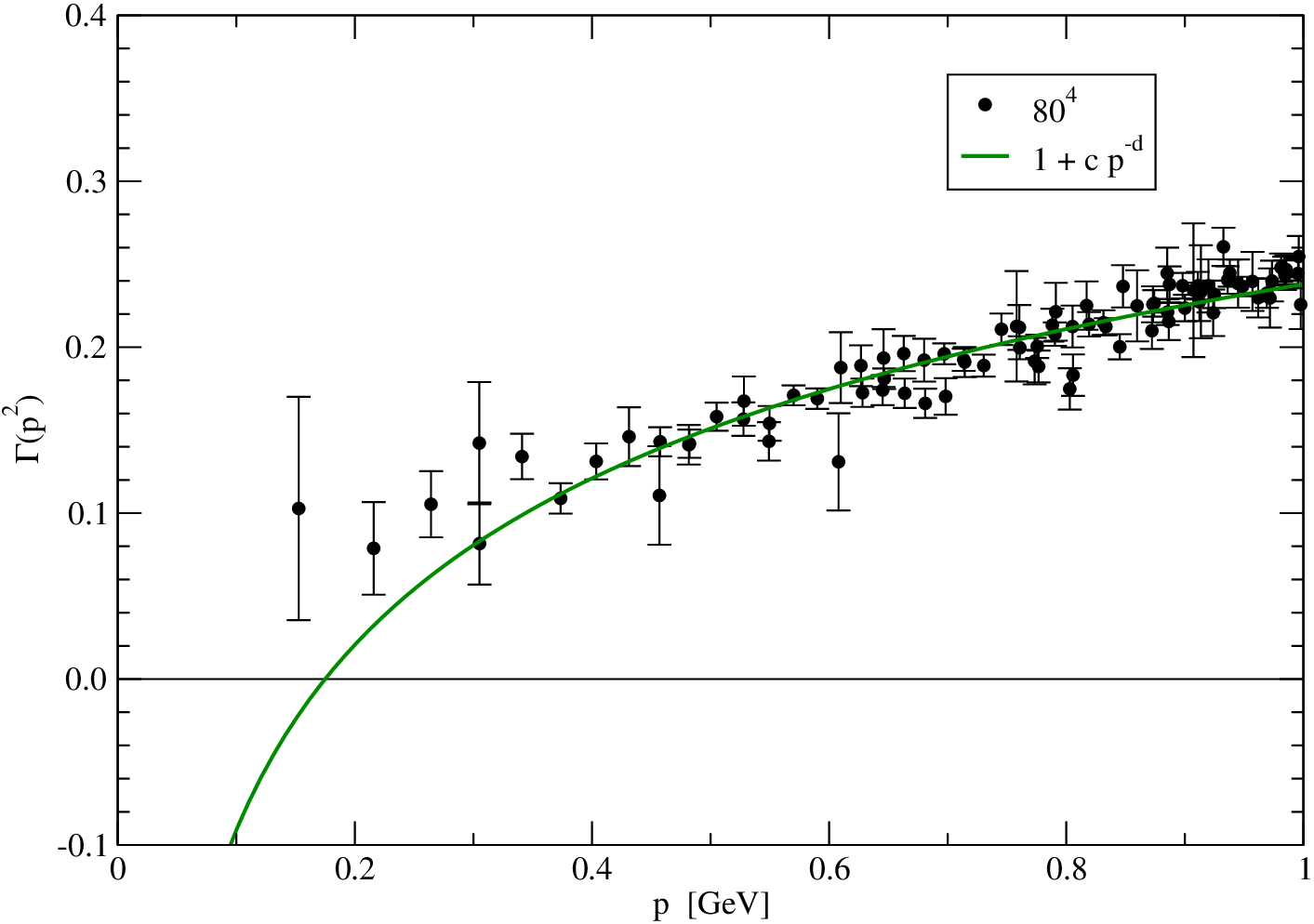}\hspace*{0.5cm}
  \includegraphics[width=0.47\textwidth]{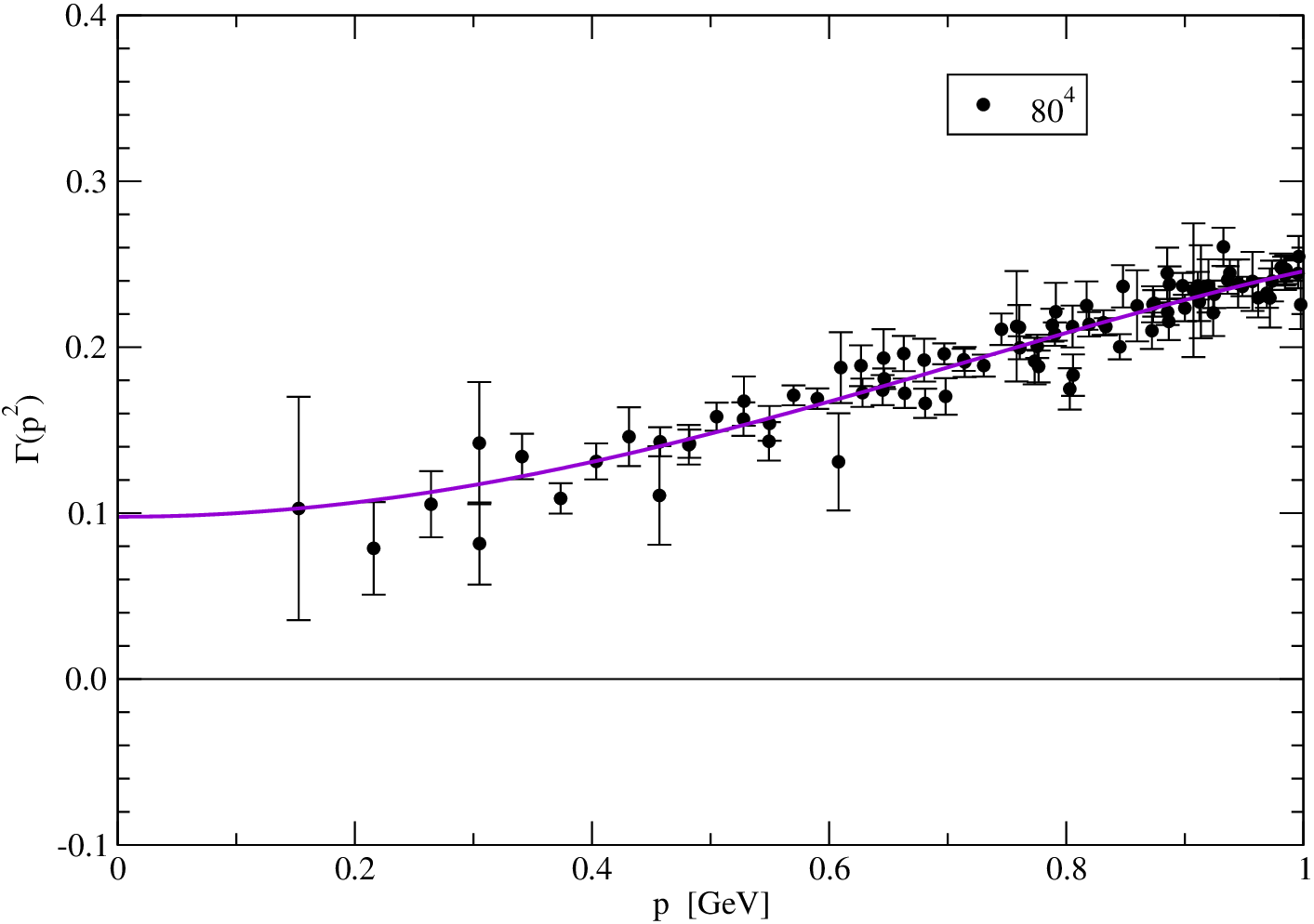} 
  \caption{Infrared $80^4$ lattice data for $\Gamma(p^2)$ together with some fitting functions. Left plot: $\Gamma (p^2) = 1 + c\,p^{-d}$; right plot: $\Gamma (p^2) = a + b p^2+c p^4$.  }
  \label{zerocrossing2}
\end{figure*}

In order to help clarifying the infrared behavior of $\Gamma (p^2)$,
we also considered a power law,  $\Gamma_3 (p^2) = 1 + c\,p^{-d}$
with the best fit resulting in  $c=-0.7621(15)$,
$d=0.1558(49)$ with a  $\chi^2/d.o.f. = 1.35$. Accordingly, the change of
sign occurs at $p_o=175$MeV; see Figure  \ref{zerocrossing2}, left
plot.
Finally we  considered the quadratic function
$\Gamma_4 (p^2) = a + b p^2+c p^4$ whose optimal parameters are  $a=0.0978(60)$,
$b=0.218(22)$, and $c=-0.070(18)$, with a $\chi^2/d.o.f. = 0.98$.
Similarly to $\Gamma_2(p^2)$, the change of sign of the form factor is also not
observed; see Figure \ref{zerocrossing2}, right plot. 

\section{Conclusions and outlook}

An improved calculation of the three gluon vertex on the lattice and
for the asymmetric momentum configuration is described.
Two different lattice volumes of $(6.5$ fm$)^4$ and $(8.2$ fm$)^4$,
with a common lattice  spacing of  $a = 0.102$ fm, have been
investigated. In general our approach to handle the fluctuations and
the lattice spacing effects seems to be able to produce compatible
smooth curves
over a wide range of momenta that extends up to $\sim$ 2.5 GeV.
Our investigation of the deep infrared region using the lattice data
and relying on fits to functional forms are inclusive in what concerns
a change of sign of $\Gamma (p^2)$.

\begin{acknowledgement}

This work was granted access to the HPC resources of
the PDC Center for High Performance Computing at the
KTH Royal Institute of Technology, Sweden, made
available within the Distributed European Computing
Initiative by the PRACE-2IP, receiving funding from the
European Communitys Seventh Framework Programme
(FP7/2007-2013) under Grant agreement no. RI-283493.
The use of Lindgren has been provided under DECI-9
project COIMBRALATT. We acknowledge that the results
of this research have been achieved using the PRACE-3IP
project (FP7 RI312763) resource Sisu based in Finland at
CSC. The use of Sisu has been provided under DECI-12
project COIMBRALATT2.
We acknowledge the Laboratory for Advanced Computing at the University
of Coimbra \cite{lca} for providing access to the HPC resource Navigator.
This work was produced with the support of INCD \cite{incd} funded by
Fundação para a Ciência e a Tecnologia, I. P. (FCT)  and
FEDER under the project 01/SAICT/2016 nº 022153.
The authors acknowledge Minho Advanced Computing Center
\cite{macc} for providing HPC resources that have contributed to
the research results reported within this paper. This work was
produced with the support of MACC and it was funded by FCT 
under the Advanced Computing Project CPCA/A2/6816/2020, platform Bob.
Work supported by national funds from FCT, within the
Projects UIDB/04564/2020, UIDP/04564/2020, and CERN/FIS-COM/0029/2017.
P. J. S. acknowledges financial support from FCT under
Contract CEECIND/00488/2017.
G. T. R. C. acknowledges financial support from FCT under Project
UIDB/04564/2020, and also from the Generalitat Valenciana
(genT program CIDEGENT/2019/040) and Ministerio de Ciencia e Innovacion
PID2020-113644GB-I00.

\end{acknowledgement}

\end{document}